\documentclass[journal=jacsat,manuscript=article]{achemso}
\setkeys{acs}{usetitle = true}
\usepackage[version=3]{mhchem} 
\usepackage[T1]{fontenc}       
\usepackage{graphicx}
\usepackage{dcolumn}
\usepackage{threeparttable}
\usepackage[usenames, dvipsnames]{color}
\usepackage{gensymb}
\usepackage{inputenc}  
\usepackage{physics}
\usepackage{bm} 
\usepackage{amsmath}  
\usepackage{amsfonts} 

\author{Laxmi Narayan Tripathi}
\affiliation[]
{Technische Physik and Wilhelm-Conrad-R\"ontgen Research Center for Complex Material Systems, Universit\"at W\"urzburg, Am Hubland, W\"urzburg-97074, Germany}
\email{ltripathi@physik.uni-wuerzburg.de}
\author{Oliver Iff}
\affiliation[]
{Technische Physik and Wilhelm-Conrad-R\"ontgen Research Center for Complex Material Systems, Universit\"at W\"urzburg, Am Hubland, W\"urzburg-97074, Germany}
\author{Simon Betzold}
\affiliation[]
{Technische Physik and Wilhelm-Conrad-R\"ontgen Research Center for Complex Material Systems, Universit\"at W\"urzburg, Am Hubland, W\"urzburg-97074, Germany}
\author{\L{}ukasz Dusanowski}
\affiliation[]
{Technische Physik and Wilhelm-Conrad-R\"ontgen Research Center for Complex Material Systems, Universit\"at W\"urzburg, Am Hubland, W\"urzburg-97074, Germany}
\author{Monika Emmerling}
\affiliation{Technische Physik and Wilhelm-Conrad-R\"ontgen Research Center for Complex Material Systems, Universit\"at W\"urzburg, Am Hubland, W\"urzburg-97074,
	Germany}
\author{Kihwan Moon}
\affiliation{Department of physics, Chung-Ang University,  Seoul, South Korea}
\author{Young Jin Lee}
\affiliation{Department of physics, Chung-Ang University,  Seoul, South Korea}
\author{Soon-Hong Kwon}
\affiliation{Department of physics, Chung-Ang University,  Seoul, South Korea}
\author{Sven H\"{o}fling}
\affiliation{Technische Physik and Wilhelm-Conrad-R\"ontgen Research Center for Complex Material Systems, Universit\"at W\"urzburg, Am Hubland, W\"urzburg-97074,
	Germany}
\alsoaffiliation[Second University]
{SUPA, School of Physics and Astronomy, University of St. Andrews,
	St. Andrews KY16 9SS, United Kingdom}
\author{Christian Schneider}
\email{christian.schneider@physik.uni-wuerzburg.de}
\affiliation{Technische Physik and Wilhelm-Conrad-R\"ontgen Research Center for Complex Material Systems, Universit\"at W\"urzburg, Am Hubland, W\"urzburg-97074,	Germany}
\title[]
{Spontaneous emission enhancement in strain-induced WSe$_{2}$ monolayer based quantum light sources on metallic surfaces}

\begin{document}
\begin{abstract}
	Atomic monolayers of transition metal dichalcogenides represent an emerging material platform for the implementation of ultra compact quantum light emitters via strain engineering. In this framework, we discuss experimental results on creation of strain induced single photon sources using a WSe$_{2} $ monolayer on a silver substrate, coated with a very thin dielectric layer. We identify quantum emitters which are formed at various locations in the sample. The emission is highly linearly polarized, stable in linewidth and decay times down to 100 ps are observed. We provide numerical calculations of our monolayer-metal device platform to assess the strength of the radiative decay rate enhancement by the presence of the plasmonic structure. We believe, that our results represent a crucial step towards the ultra-compact integration of high performance single photon sources in nanoplasmonic devices and circuits.
\end{abstract}
\newline
\maketitle

Single photon sources are considered as a key building block for quantum networks, quantum communications and optical quantum information processing \cite{Pan2012,Kok2007,OBrien2007,Kim1999,Kremer2014}. To fully harness the properties of such non-classical light sources, core requirements include their long-term stability \cite{Kurtsiefer2000, Michler2000}, brightness\cite{Santori2007, Ding2016, Unsleber2015,Somaschi2016}  and scalability in the fabrication process. Recently, quantum light emission from inorganic two dimensional layers of transition metal dichalcogenides (TMDC)\cite{Srivastava2015,Chakraborty2015,He2015,Koperski2015,He2016a} has been demonstrated. While the nanoscopic origin of tight exciton localization is still to be explored, engineering the morphology of carrier substrates and thus the strain field in the monolayers \cite{Kumar2015a} has enabled position control over such quantum emitters \cite{Kern2015,Palacios-Berraquero2017,Branny2017}.

One outstanding problem, which we address in this report, is the emission enhancement of such quantum emitters in atomic monolayers. The layered nature of the materials and their intrinsic robustness with regard to open surfaces (due to absence of dangling bonds) naturally puts plasmonic approaches in the focus of interest. A single dipole emitter close to a plasmonic nanoparticle, which act as an optical antenna\cite{Novotny2011,Biagioni2012}, experience a modified photonic mode density, leading to enhanced radiative decay rates and thus a spontaneous emission enhancement. The enhanced  intensity  results from an amplified electric field intensity due to localized surface plasmon resonance of metal nanoparticles \cite{Haridas2011a,Eggleston2015,Kern2015}.Plasmonic tuning of the optical properties of molecules, such as dyes close to a metal surface is a topic which is subject to investigations since the 1980s. Pronounced coupling phenomena of dye molecules with  surface plasmon resonances in ultra-thin silver films has been shown via luminescence and absorption studies\cite{A.M.Glass1980}, as well as resonant transmission \cite{Garoff1981}. Enhanced fluorescence of molecules coupled to Ag-islands has been studied in \cite{Weitz1982}, whereas fluorescence quenching of dye molecules or colloidal CdSe quantum dots in the closest vicinity of metallic surfaces has also been identified to act significantly on the emitters' decay dynamics \cite{Dulkeith2002,Tripathi2015}. Hence it is important to separate the emitters from the metallic layers via a non-conducting spacer (For example, $ Al_{2}O_3 $) of few nanometers thickness. Furthermore, plasmonic nanoparticles close to emitters do not only enhance the emission rate via the Purcell effect but also can yield enhanced excitation rates. In fact the over-all PL enhancement of emitters very close (few nm) to plasmonic nanoparticles is a result of  both excitation and emission rates enhancements \cite{Khurgin2009}.  

Recently, the deterministic coupling of single quantum emitters in thin hBN layers to plasmonic nanoparticle arrays has been shown \cite{Tran2017}, and even in comparatively simple schemes, near unity absorption of light has been demonstrated by coupling of optical modes at an interface between a WSe$_2$ layer and silver film. This indicates a strongly enhanced local density of states in such architectures \cite{Jariwala2016}. Enhancement of photoluminescence (PL) has also been  observed from atomic monolayer materials on metallic layers \cite{Cheng2017,Zhou2017,Loh2017}. Metallic substrates have the advantage of reflecting almost all of the photons towards the collecting objective, while the transmission or absorption of single photons inside the substrate is the main source of loss of the single photons. In addition, metallic nano-particles on a  metal surface act as native antennas \cite{Giannini2011,Novotny2011,Anger2006,Koenderink2017}, which in principle allow to engineer the spontaneous emission rates of surface-near quantum emitters \cite{Lapin2015,Anderson2006,Crozier2003,Novotny2006,Novotny2010,Anger2006,Bozhevolnyi2016,Bozhevolnyi2017}. 

In this report, we demonstrate strain induced creation of single photon sources in a monolayer of WSe$_{2}$ in the close proximity of a nano-modulated metallic substrate. Our source is temporarily stable and exhibits robust emission features comparable to monolayer single photon sources on dielectric substrates. Our joint experimental-theoretical investigation suggests, that the spontaneous emission of our single photon emitters is enhanced by coupling to plasmonic modes in the metal. 
\begin{figure}
	\centering
	\includegraphics[scale=0.7]{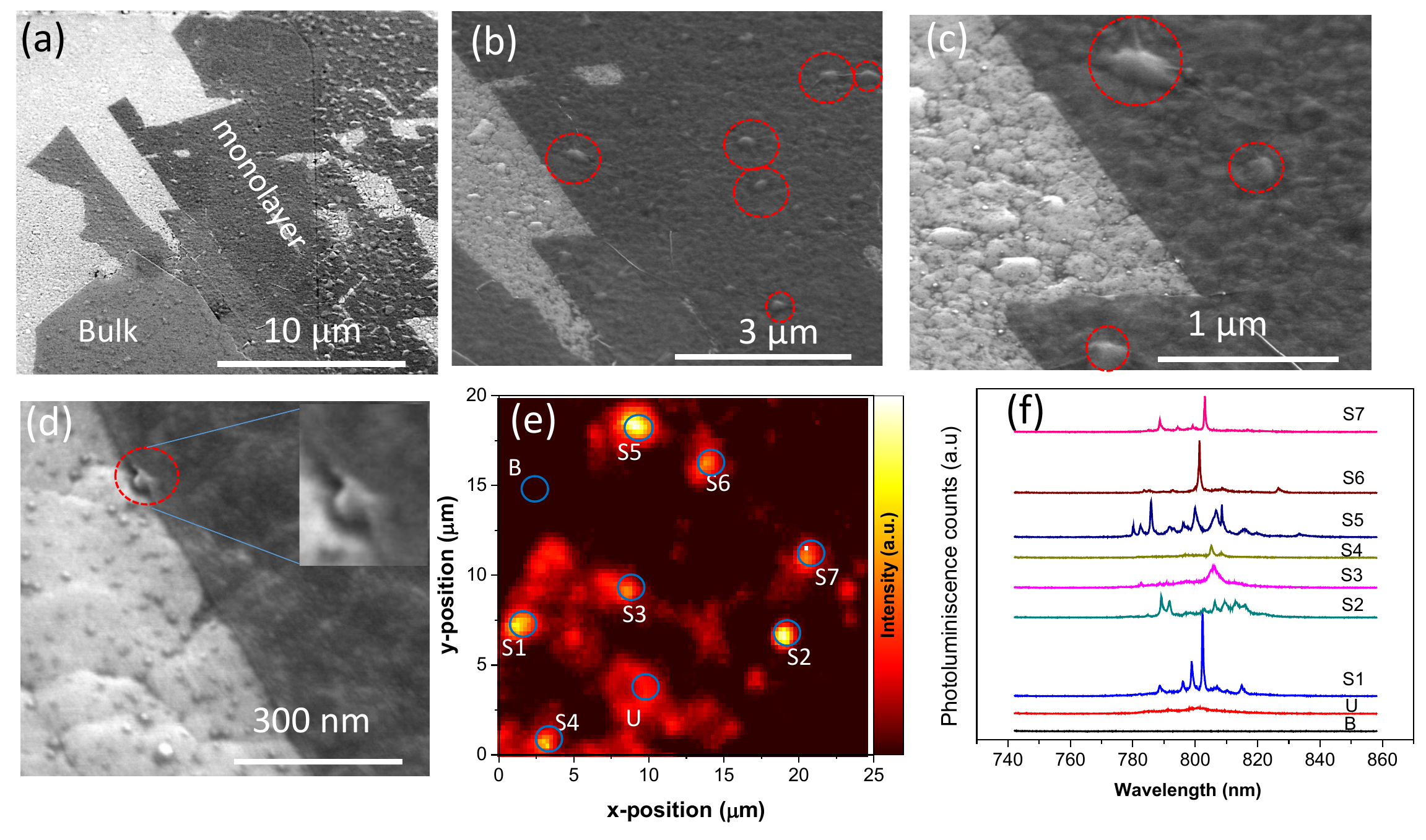}
	\caption[]{\label{Sample_PLchararateristics}(a) Bright field scanning electron microscope (SEM) image of the monolayer and bulk WSe$_{2}$ crystal. White small features are silver islands and nano-particles protected with 3 nm $ Al_2O_3 $ layer. (b-d) High resolution SEM images showing conformal morphologies of WSe$_{2}$ and silver surface. Red dotted circles show strained monolayer around such nanoparticles. d) A monolayer is seen to bend over a silver nanoparticle (dotted red circle and inset: zoomed in picture of the strained monolayer due to the nanoparticle). (e) Spatially resolved and wavelength integrated (750-850 nm) PL intensity map of the structure, depicting bright emission spots (encircled) occurring on the monolayer. f) A series of representative  photoluminescence spectra  picked from the bright spots on the sample, corresponding with positions S1-S7.,  U; homogeneous background (due to unstrained monolayer), B; blank area devoid of monolayer.}
\end{figure}
The base structure of our study consists of a  600 $\mu$m sapphire substrate, which is capped by 10 nm of chromium and 200 nm of silver. Fig.~1(a-d) depicts scanning electron microscope images of a mechanically exfoliated  WSe$_{2}$ monolayer transferred onto the silver substrate. In particular from the close-up images in Fig.~1(b-d), it becomes evident that the monolayer bends around exposed regions of the metal surface, and fully adapts its morphology. In addition, we observe the formation of nano-wrinkles, which possible yield nanoscopic strain fields in the monolayer. 

In order to test the influence of the macroscopic roughness of our prepared surface on the emergence of strongly localized excitons in the TMDC monolayers, we studied our sample via position resolved scanning PL with a step size of 250 nm. The wavelength-integrated corresponding PL intensity map is depicted in Fig.~1e. While, surprisingly, we did not see any significant signal from the attributed free exciton or trion resonance in WSe$_2$, we observe a strongly inhomogeneous emission pattern giving rise to distinct high intensity spots in a wavelength range between 780 nm and 840 nm. In order to shed more light into the origin of these spots, we plot the selected spectra corresponding to positions S1-S7 in Fig.~1f. Here, it becomes evident, that a large fraction of these sites are affiliated with the emergence of sharp emission lines, which are prototypical for strongly localized quantum emitters in atomic monolayers. The absence of the distinct free exciton resonances could be a result of exciton diffusion towards such distinct localization sites, which form an attractive potential for excitons. We believe, that the bended monolayer effectively leads to an exciton funneling effect, which particularly exposes some crystal defects and nanowrinkles as emitters of narrow band spectral features.

In order to analyze the optical quality of our emitters, we study the linewidth of each emerging peak in our monolayer. To reduce significant power broadening and laser induced heating, the power was reduced to 708 $ W/cm^{2} $ and the spectrum was integrated for 10 sec. The acquired linewidth statistics is shown in Fig.~2a. Although we identify emission features scattered over a wide range of linewidths (few 100 $\mu$eV to few meV), we observe various  emitters close to the resolution limit of our optical spectrometer (75 $\mu eV$, see inset of Fig.~2a). We plot the emission intensity (Fig.~2b) and linewidth (Fig.~2c) of a selected emitter as a function of the excitation power of the driving laser.
The excitation power dependent photoluminescence intensity (Fig.~2b) can be approximated by an equation for a two level system\cite{Kumar2015a},  I = $ I_{sat} (1/(1+P_n/P_{exc}))$ where, $ I_{sat} $ is the saturation intensity and $P_{exc}$ is excitation power. $ P_{n} $ is a fit parameter normalizing the excitation power.  

\begin{figure}
	\centering
	\includegraphics[scale=0.5]{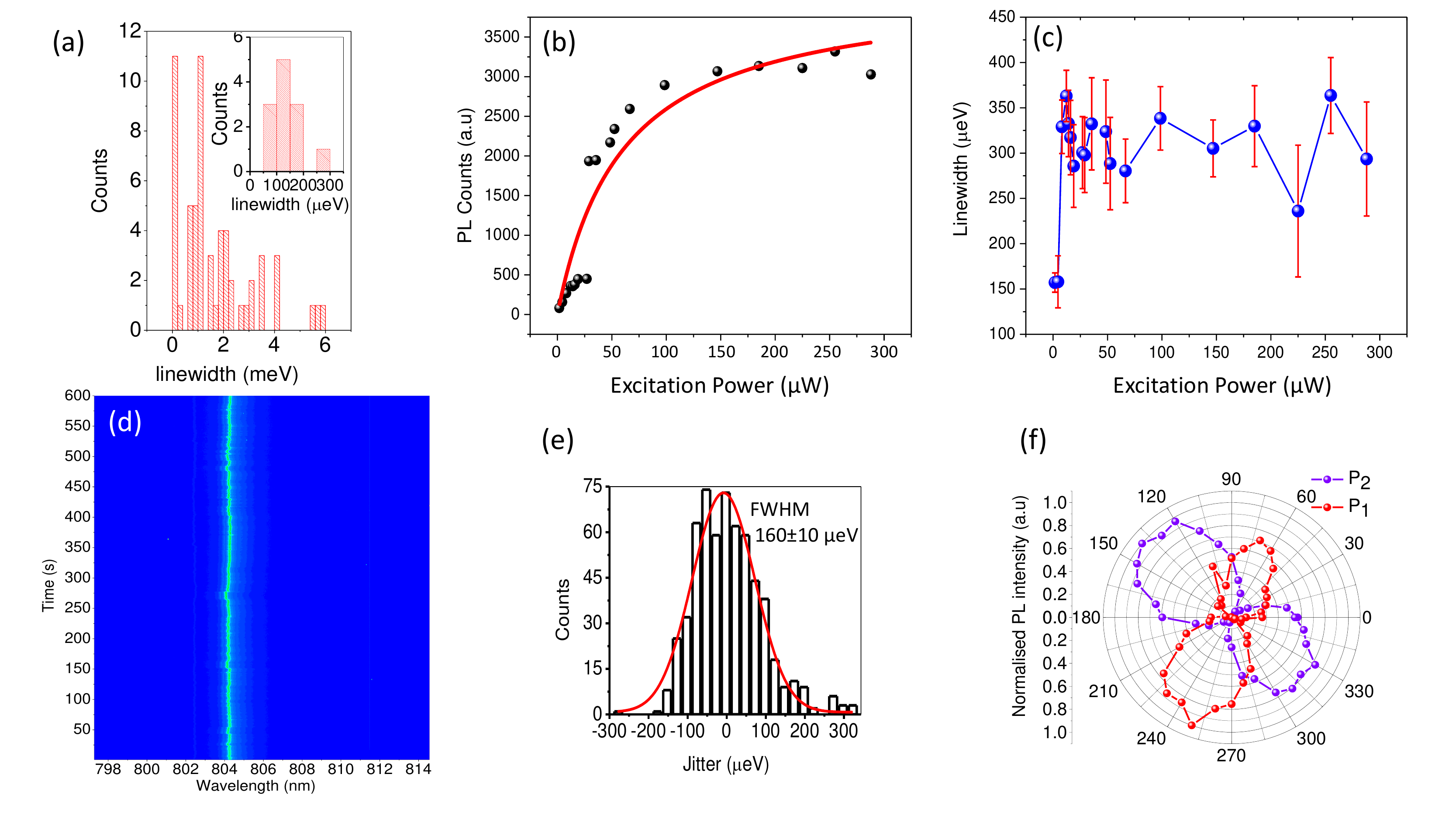}
	\caption[]{\label{linewidty-Power} a) Linewidth statistics of  various emitters measured at  power of 50 $\mu$W $ \approx 0.8~P_n $. b) Photoluminescence intensity saturation curve with increasing excitation power of a CW 532 nm laser. The fit to the data is given by an equation known for a two level system. c) Linewidth of the emitter studied in panel (b) as a function of the excitation power. d) Photoluminescence intensities map of an emitter over span of 600 s recorded at an interval of 1 s. e) Histogram of spectral jitter of the emitter whose PL intensities with time is shown in (d). f) Polar plot of the normalized PL intensity of two selected emitters $ P_1 $ and $ P_2 $ as function of the detected polarization angle. All the measurements are done at 5 K.}
\end{figure}

 The excitation power dependent linewidth (Fig.~2c) is stable against increasing  excitation power and signifies the robustness against severe power broadening. Spectral wandering of the emitter over time,  a characteristic feature of a single quantum dot in unsteady charge environment due to  spectral diffusion\cite{Seufert2000} and charge fluctuation\cite{Pietka2013}, was measured for 600 sec for one selected QD in Fig.~2d. Here, we clearly observe a  spectral wandering on a macroscopic time scale, which further contributes to the linewidth broadening analyzed in Fig.~2a. The map suggests that the emitter does not blink on the time scale of seconds. A histogram of spectral jitters over 600 sec of the same emitter,  with 1s time bin,  is shown in Fig.~2e. A Gaussian fit of the histogram gives FWHM of $160\pm10~\mu eV$.

In order to assess the  polarization properties of the emitters, we measured a series of polarization dependent PL spectra using a half wave plate and a linear polarizer. The resulting  polar plot of two of such emitters, called  $ P_1 $ (786.2 nm) and $ P_2 $ (791.5 nm) as a function of the detected polarization angle is shown in Fig.~2f. We can see that the emitters are highly linearly polarized (with a degree of polarization (DOP) of 92 \% for $P_1$ and 94\% for $P_2$), and differ in orientation by an angle of $ \approx 80 \degree $. One plausible origin of the orientation and high directionality could be an effect of light matter coupling of the emitters and silver nanoparticles. Different orientation of adjacent emitters further suggest that emitters $ P_1$ and $ P_2$ might be located at strained edges around the same silver nanoparticle. However, we also note that high degrees of linear polarization have been reported on for emitters in WSe$_2$ on dielectric substrates, which has been associated with anistotropic strain in the crystal \cite{Kern2015}. 

 \begin{figure} 
 	\centering
 	\includegraphics[scale=0.65]{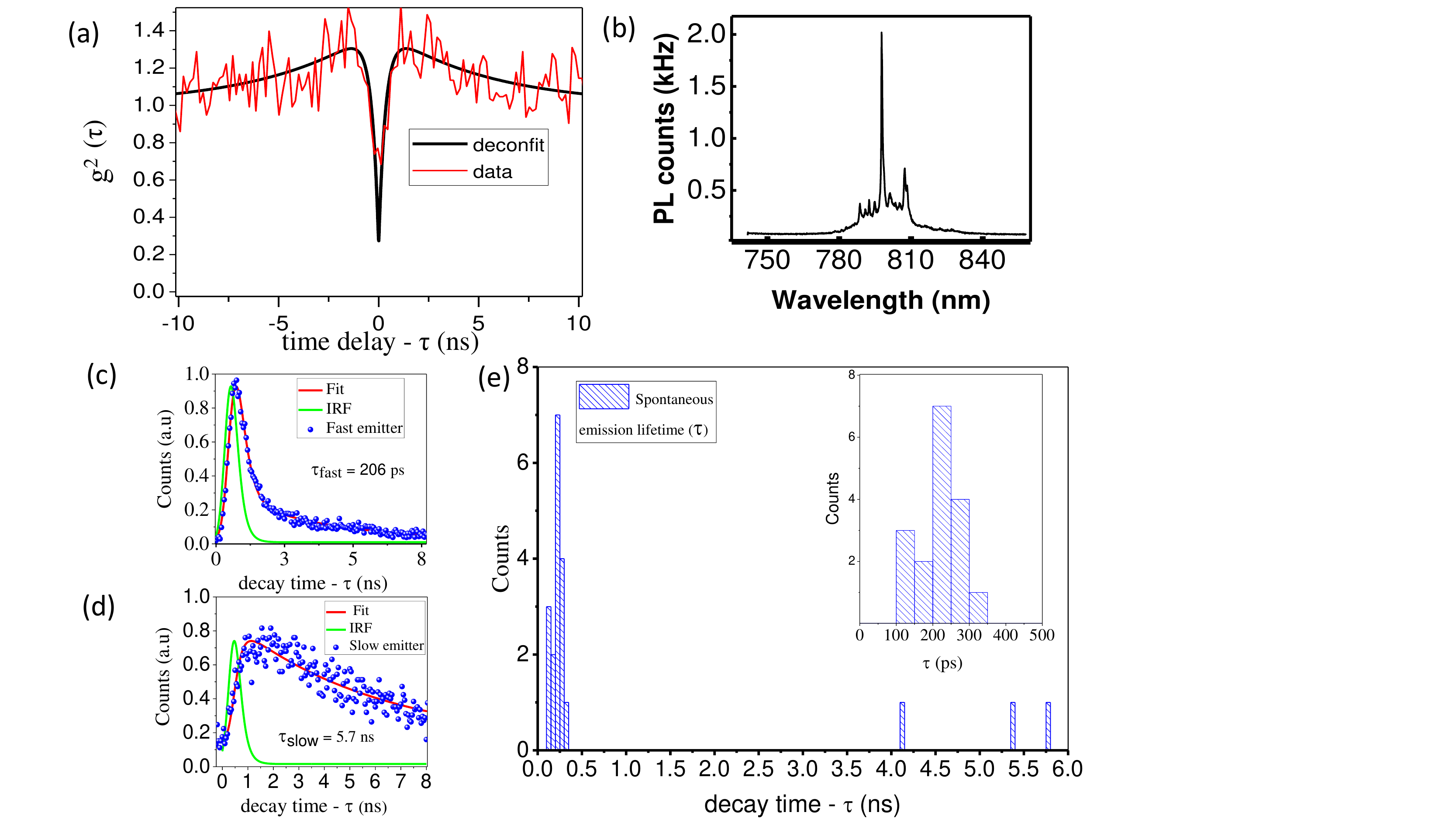}
 	\caption{\label{correlation and lifetime}a), The second order correlation function,  $g^{2}(\tau)$  as function of time delay($ \tau $). The fit after deconvolution (deconfit) showing the second order correlation function at zero delay $g^{2}(0)$ to be 0.25.  b),  Full PL spectrum selected for the $g^{2}(\tau)$ measurement. The selected  peak is indicated by a blue arrow.The time-resolved decay dynamics (blue dots), Instrument response curve (green line), double exponential fit (red curve) for  (c), a fast emitter  and  (d), a slow emitter. (e),  Histogram of spontaneous emission lifetimes$ (\tau $), extracted from  exponential fits of time-resolved spectra of 20 quantum emitters as shown in fig(c) and fig (d). Inset: histogram of spontaneous emission lifetime distributions of fast emitters.}  
 \end{figure}

To test our system with regard to the statistics of emitted photons, we performed second order correlation  measurements by exciting the sample with a 532 nm CW laser (Fig.~3a). We selected a dominant emission feature (shown in Fig.~3b), which was spectrally filtered (bandwidth: $\approx $ 300 $\mu$eV, 300 grooves/mm grating) and passed it to a fiber coupled Hanbury Brown and Twiss (HBT) interferometer. We observed a sharp and narrow anti-bunching feature at zero delay times $ (\tau = 0 $). In addition, the correlation exhibits a slight bunching behavior at finite delay times. The experimental data, as well as a fit to the data are depicted in Fig.~3a. There, we explicitly accounted for the finite time resolution of our single photon detectors (350 ps) by introducing a Gaussian response function $ g_{d} $\cite{He2016a}.  
The overall fitting function thus writes  
$ g^{2} (\tau) = g_{d} \times \left(1 - (a \cdot e^ -\frac{\abs{ \tau}}{\tau_{c_1}}) +(1-g^{2}(0)-a) e^ -\frac{\abs{ \tau}}{\tau_{c_2}}\right)$.
There, $  \tau_{c_1} $ (5.3 $\pm $ 0.8 ns) is the characteristic timescale for the observed bunching, $ \tau_{c_2} $ (0.3 $ \pm  $ 0.1 ns) is the time scale of the observed antibunching and \textit{a} the ratio between both. Since the measurement has been carried out below the saturation level of the emitter, the timescale of the antibunching signal can be can be directly related to the lifetime of the studied signal. We note, that typical radiative decay times of localized quantum emitters in  $ W Se_{2} $ on dielectric substrates have been reported to be on the order of several nanoseconds (see table 1).

\begin{table}[h]
	\caption{Literature reports of  decay times of localized quantum emitters in $ WSe_{2} $ on dielectric substrates.}
	\begin{center}
		\begin{threeparttable}
			\begin{tabular}{ |c|c|c|}
				\hline
				\textbf{Reference}        & \textbf{Decay time (ns)} &       \textbf{Substrates}        \\ \hline
				Ref.\cite{Kumar2015a}     &           4.1           &      bare silicon\tnote{1}       \\
				Ref.\cite{He2015}         &           3.6           &  \tnote{2} SiO$_2$   on silicon  \\
				Ref.\cite{Srivastava2015} &        (1.5-2.5)        &        SiO$_2$ on silicon        \\
				Ref.\cite{Branny2017}     &           2.8           & Polymer pillars array on silicon \\
				Ref.\cite{Tonndorf2015}   &           1.8           &        SiO$_2$ on silicon        \\
				Ref.\cite{Amani2016}      &      \tnote{3} 19.3      &        SiO$_2$ on silicon        \\
				Ref.\cite{Ye2017}         &    $ 10.3 \pm 0.5  $    &  SiO$_2$ on silicon  \tnote{4}   \\
				Ref.\cite{Palacios-Berraquero2017}  &  3-8          &  silica nano-pillars on silicon \\  \hline
			\end{tabular}

			\begin{tablenotes}
				\item[1] Etched hole arrays on bare silicon substrates.
				\item[2] $ SiO_2 $; silicon dioxide. 
				\item[3] Decay time of a quantum emitter measured using quantum yield.
				\item[4] In situ strain tuning due to applied pressure.
			\end{tablenotes}
			
		\end{threeparttable}
	\end{center}
	\label{table:decaytimes_defects}
\end{table}
The TMDC-metal hybrid is quite stable against atmospheric (air) and temperature conditions, allowing us to measure the localized emission spectra even after repeated thermocycles (cryostat-temperature, He flow at 5 K to 297 K) and laser intensities, without damaging the strained monolayer for nine months.  

The extracted emitter lifetime of only 300 ps suggests that the spontaneous recombination rate in our quantum emitter is subject to some enhancement by coupling to plasmonic modes in the silver layer. The observed bunching at much longer characteristic time scales is indicative that longer lived metastable states also play a role in our system. We believe, that the contribution of a localized dark exciton, which was recently observed in $WSe_2$ based quantum emitters, could yield such a behavior \cite{He2016}. In addition, we furthermore believe that the presence of such long lived characteristic timescales in our photon statistics excludes very strong non-radiative recombination as a sole source of significant lifetime shortening in our experiment. The extracted second order correlation function at zero delay, $ g^{2}(0)$ is 0.25, adds evidence that our system acts as a single photon source. 

In order to confirm the fast recombination of our quantum emitters, we additionally performed time correlated single photons measurements of 20 emitters at different positions of the monolayer. Since the time resolution of the single photon detector is on the same order of magnitude as the lifetime of the emitters, the measured decay M($\tau$) is determined both by the instrumental response function IRF($\tau$) and the time-resolved emission of the sample E($\tau$). Therefore each measured decay was fitted by iterative re-convolution of the IRF with a bi-exponential decay model $E(\tau)=a \cdot e^{-\frac{\tau-\tau_0}{\tau_1}}\cdot \Theta(\tau-\tau_0)+(1-a) \cdot e^{-\frac{\tau-\tau_0}{\tau_2}}\cdot \Theta(\tau-\tau_0)$, where $\Theta$ describes the Heaviside step function. The instrument response was experimentally extracted by  measuring the reflected laser signal and for the re-convolution it was fitted with an exponentially modified Gaussian function. Indeed, we observed two different classes of quantum emitters in our study , which are exemplary depicted in Fig 3c) and 3d). For the class of emitters corresponding with the time-trace in Fig 3c), we observe a very fast emission decay between 100- 500 ps, being significantly faster than reported values in the same material system on dielectric surfaces (see table 1). The statistics of the recombination time from these emitters is plotted in the inset of Fig 3e). The second class of emitters in our study yields decay times on the order of various nanoseconds (Fig 3d). We believe, that these emitters emerge from wrinkles in our strained monolayers which do not couple to the silver surface.

In order to add further confidence, that the origin of the enhanced  radiative decay rates is a result of coupling of our quantum emitters to plasmonic resonances evolving in the silver surface in the close proximity \cite{Jariwala2016}, we calculated the spontaneous emission (SE) decay rate enhancement by finite-difference time-domain (FDTD) analysis of a dipole emitter near a cone-shaped a metallic nanostructure. Comparatively, we study a dipole emitter near sapphire nanocones or flat substrates. Since the quantum electrodynamic treatments of SE leads proportional results to classical treatments of SE, the SE rate is related to the classical dipole emission power in air and the nanostructures \cite{Englund2005, Xu1999}. 

\begin{figure} 
	\centering
	\includegraphics[scale=0.5]{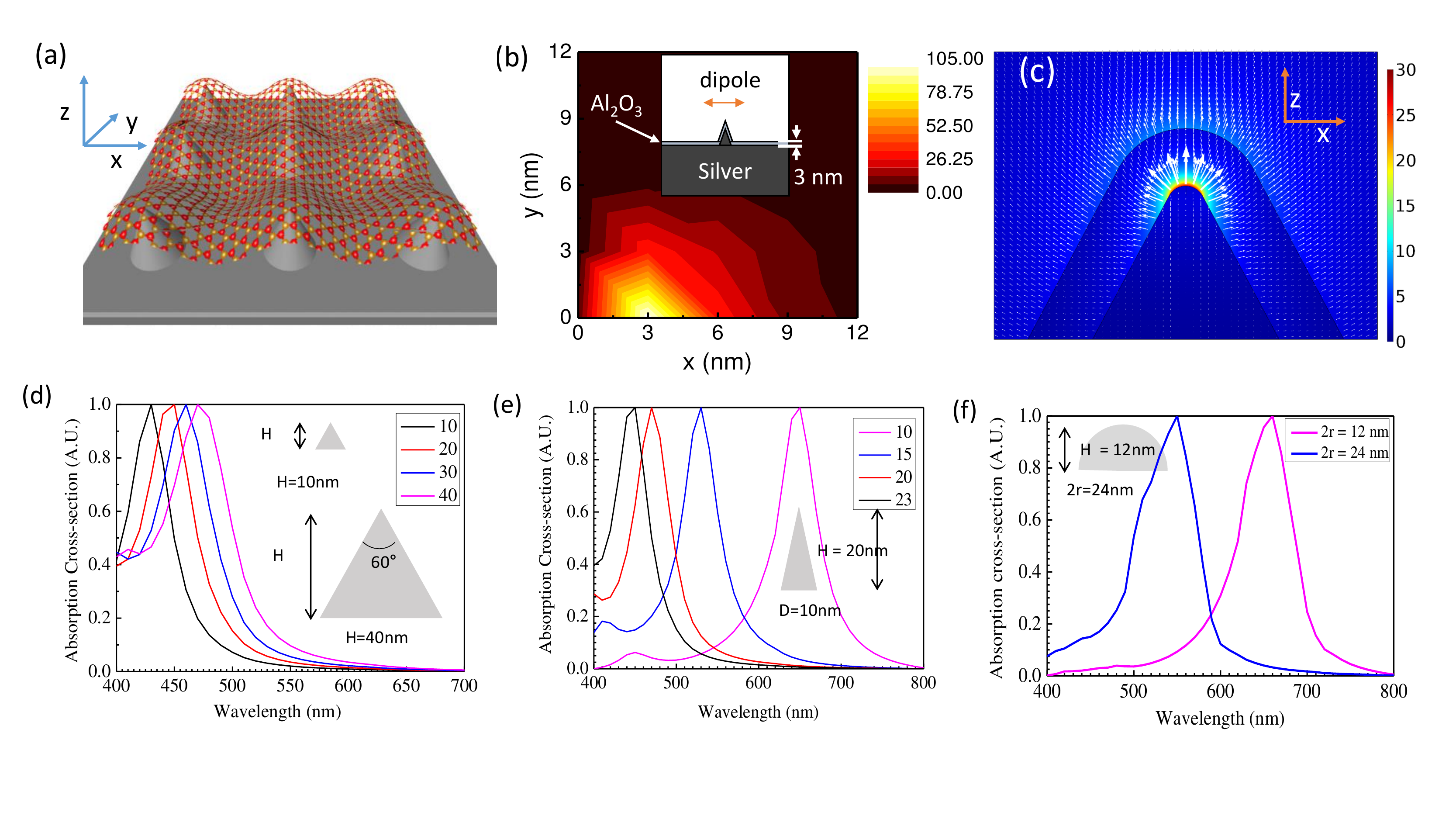}
	\caption{FDTD simulation: a) Artistic sketch of the assumed sample configuration, featuring the monolayer wrapped over metal cones evolving on the silver-Al$_{2} O_{3} $ surface. b) Decay rate enhancement of a horizontal dipole over a silver nanocone. c) Vector map of the field distribution and electric field enhancement E/$E_0$  around a silver nanocone excited by an electromagnetic field $E_0$ of 1 V/m.  Calculated spectra of different nanostructures: d) nanocones for a fixed aspect ratio with all apex of an angle of 60\degree, from left to right: height from 10 nm to 40 nm,  e) nanocones for a fixed height of 20 nm, from left to right: base diameter from 23 nm to 10 nm, f) half-spheres for a fixed height of 12 nm with different base diameters, 24 nm (blue) and 12 nm (magenta).}
\end{figure}

A sketch of the assumed scenario is shown in Fig.~4a. A horizontal dipole emitter is placed on top of the nanocone structure with a wavelength of 800 nm and we calculated the output power in the nanostructure, as we move the position horizontally with a spatial resolution of 3 nm. The reference output power of the dipole is calculated by putting the dipole in the background of air. The grid resolution is 1 nm.  The decay rate enhancement map for a dipole emitter around a silver nanocone is shown in Fig.~4b. We observe, that nearly 100 times decay rate enhancement is expected for a dipole, which is located 3 nm away from the tip of the cone, while a negligibly small value is observed for a non-metallic reference structure purely comprising a sapphire cone (not shown). The vector field distribution and the field enhancement E/$E_0$ for our metallic, $Al_2O_3$ coated device is shown in Fig.~4c, which is calculated by the finite element method based software COMSOL Multiphysics. The calculation around a silver cone shows that the field is strongly enhanced 30 times compared with an incident field  which supports the large decay rate enhancements. The calculation  around a silver cone furthermore shows that the field is perpendicular at the tip, hence prohibiting the coupling to a horizontally oriented dipole. However, coupling between the horizontal component of the field at the edge of the cone and the exciton dipole in the monolayer lead to significant radiative decay rate enhancement. Here, we would like to state, that there is a persisting likelihood that the localized dipole in WSe$_2$ also features a vertical component, which would even further increase the coupling strength in our metal cone structures.

Figure 4d) depicts resonance spectra for nano-particles with a cone angle of $60^\circ$ as the height of the cone becomes smaller (i.e. for flat nanostructures). As expected, the absorption spectrum is subject to a blue shift under out of plane (parallel  to the cone axis) polarization. This result is consistent with previous studies \cite{Schaefer2013}. In contrast, if the height of the cone is fixed and the base diameter of the cone varies, the spectra show a distinct red-shift  with decreasing diameter. Indeed, in Fig 4e, the absorption spectrum of a nanocone with 10 nm height is centered around  650 nm. Lastly, in Fig 4f, we plot absorption spectra for hemispherical silver nanoparticles, with a fixed height of 12 nm and a variation of the base diameter (12 nm vs. 24 nm). In fact, these sizes match the approximate structure dimensions of the smallest class of our metallic particles (e.g. depicted in Fig 1d). We note, that considering the dielectric environment due to WSe$_{2}$ on the nanostructure which is known to induce a further redshift on the spectrum of several 10~nm \cite{Krasnok2018,Abid2016}, the smaller silver nanoparticles (diameter: 12 nm) can be expected to approach the resonance frequency range (720-800 nm) of our quantum emitters. Furthermore, it is indeed known that the spontaneous emission enhancement of a plasmonic device can be the strongest on the low energy side of the actual plasmonic resonance \cite{Thomas2004,Anger2006,Bharadwaj2007}.

In conclusion, we demonstrated the formation of quantum dot-like structures in a monolayer of WSe$_2$ exfoliated onto a rough metallic surface. The metallic surface serves a dual purpose of not only creating quantum emitters in the monolayer via nanoscopic strain engineering, but also by enhancing the spontaneous emission rate. The latter is experimentally confirmed by time resolved studies, yielding decay times down to 100 ps. Our numerical calculation suggests decay rate enhancements due to a nanocone antenna. We verify that the coupling efficiency is optimum near the edges of the cone, rather than the center of the tip. Our work on the hybrid device will motivate the further research on integration of TMDC-based single photon sources and plasmonic waveguide/nano-circuits for quantum photonics/plasmonics application on a fully metallic substrate.

\textbf{Methods:} 
\textit{Sample fabrication:} Silver substrates were prepared by deposition of $ 200\pm10 $ nm silver thin film on a sapphire substrate using e-beam evaporation. The silver surface was polished by  Ar ion milling  at angle 75$\degree$ for 4 min 19 second and then at 85$\degree$ for 4 min 35 second. To protect the surface from oxidation, 3 nm  $ Al_{2}O_{3} $ were deposited via atomic layer deposition (ALD) \cite{Chen2013d}. Next, monolayers of WSe$_2$ were mechanically exfoliated \cite{Castellanos-Gomez2014} via commercial adhesive tape from bulk WSe$_{2}$ crystals (2D semiconductors, USA) onto a PDMS stamp, and subsequently transferred  onto the silver/insulator surface.

\textit{Optical Spectroscopy:} Spatially and spectrally resolved  photoluminescence (PL) measurements, which were carried out on  the sample attached to liquid helium flow cryostat at temperature of 5K. We excited the sample with a 532 nm CW laser and collected the PL-signal with a Mitotoyo objective (50x and 0.42 NA). The excitation laser was filtered by a high pass  color filter (630 nm). The signal is then spectrally filtered by selecting 1500 g/mm for higher resolution  and 300 g/mm  for capturing full spectrum, using Princeton Instruments, SP2750i spectrometer consisting of  a liquid nitrogen cooled Charged Coupled Device (CCD). For time resolved photoluminescence measurement, the sample was excited with 445 nm pulsed laser with 76 MHz repetition rate, 12 $\mu$W excitation power throughout the experiments. The time resolved spectrum was recorded with an avalanche photo diode (APD) having a timing resolution of 350 ps. The second order correlation measurement (HBT interferometry)  was done by including  another APD (Count T100-FC). A fiber beam splitter was used to connect the two APDs after filtering  the signal via the  monochromator. The linewidth of an emitter is obtained by a Lorentzian fit of spectrum obtained after subtraction of baseline.

\textit{ Numerical Simulation:} A  nanocone with a height of 20 nm and a diameter of 23 nm is assumed to mimic the fabricated rough metallic surface. Al$_{2} O_{3} $ with the refractive index of 1.671 is conformably coated with a thickness of 3 nm over whole nanostructure, where the dipole emitter is assumed to be placed on the $Al_2O_3$ layer likewise the TMDC layer in the experiment. The material of dielectric substrate, sapphire has the index of 1.7522 and no absorption for the target wavelength of 800 nm. Optical constant of silver is modeled based on the experimentally determined dielectric function of silver \cite{JOHNSON1972}. For the Finite Element Method (FEM) simulation the tip and the edge of the nanocone is modeled with a radius of 1 nm.

\textbf{Author Contributions:}

C.S and S.H initiated and guided the work, L.T fabricated the sample, L.T, O.I, M.E and L.D were carrying out experiments, S.-H.K, Y.J.L and K.M performed simulation. L.T, O.I, S.B, C.S analysed and interpreted the data. L.T, O.I, S.B, S.K and C.S wrote the manuscript with input from all authors.

\textbf{Acknowledgment:}

 We thank Prof Dai Sik Kim, Seoul National  University, Seoul, South Korea for preparation of the silver substrates. We also thank Tristan Harder for helping with a matlab code used in optical analysis. Our sincere thanks  for funding  are due to  State of Bavaria; H2020 European Research Council (ERC) (Project Unlimit-2D).
 
\providecommand{\latin}[1]{#1}
\makeatletter
\providecommand{\doi}
{\begingroup\let\do\@makeother\dospecials
	\catcode`\{=1 \catcode`\}=2 \doi@aux}
\providecommand{\doi@aux}[1]{\endgroup\texttt{#1}}
\makeatother
\providecommand*\mcitethebibliography{\thebibliography}
\csname @ifundefined\endcsname{endmcitethebibliography}
{\let\endmcitethebibliography\endthebibliography}{}

\end{document}